\documentclass[10pt,letterpaper]{article}
\usepackage[top=0.85in,left=2.75in,footskip=0.75in,marginparwidth=2in]{geometry}

\usepackage[utf8]{inputenc}

\usepackage{cite}

\usepackage{nameref,hyperref}

\usepackage{microtype}
\DisableLigatures[f]{encoding = *, family = * }

\raggedright
\usepackage{changepage}

\usepackage[aboveskip=1pt,labelfont=bf,labelsep=period,singlelinecheck=off]{caption}

\makeatletter
\renewcommand{\@biblabel}[1]{\quad#1.}
\makeatother

\usepackage{lastpage,fancyhdr,graphicx}
\usepackage{epstopdf}
\fancyheadoffset[L]{2.25in}
\fancyfootoffset[L]{2.25in}

\usepackage{color}

\definecolor{Gray}{gray}{.25}

\usepackage{graphicx}

\usepackage{sidecap}

\begin{document}
\vspace*{0.35in}

\begin{flushleft}
{\Large
\textbf\newline{A single beam Cs-Ne SERF magnetometer with differential laser power noise suppression method}
}
\newline
\\
Yao Chen\textsuperscript{1},
Yintao Ma\textsuperscript{1},
Minzhi Yu\textsuperscript{1},
Ning Zhang\textsuperscript{2,*},
Libo Zhao\textsuperscript{2},
Xiangguang Han\textsuperscript{2},
Zhuangde Jiang\textsuperscript{1}
\\
\bigskip
\bf{1} State Key Laboratory for Manufacturing Systems Engineering and the International Joint Laboratory for Micro/Nano Manufacturing and Measurement Technologies, School of Mechanical Engineering, Xi’an Jiaotong University, Xi’an 710049, China.
\\
\bf{2} Research Center for Quantum Sensing, Intelligent Perception Research Institute, Zhejiang Lab, Hangzhou 310000, China
\\
\bigskip
* ningzhang@zhejianglab.com

\end{flushleft}

\section*{Abstract}
We describe a single beam compact Spin Exchange Relaxation Free(SERF) magnetometer whose configuration is compatible with the silicon-glass bonding micro-machining method. A cylindrical vapor cell with 3mm diameter and 3mm in length is utilized in the magnetometer. In order to reduce the wall relaxation which could not be neglected in micro-machined SERF magnetometer, 3 Amagats(1Amagat=2.69$\times$ 10$^{19}$/cm$^3$) neon buffer gas is filled in the vapor cell and this is the first demonstration of a Cs-Ne SERF magnetometer. We also did a simulation to show that neon is a better buffer gas than nitrogen and helium which is typical utilized in vapor cells. In order to reduce the laser amplitude noise and the large background detection offset which is reported to be the main noise source of a single beam absorption SERF magnetometer, we developed a laser power differential method and a factor of 2 improvement of the power noise suppression has been demonstrated. Finally, we did an optimization of the magnetometer and sensitivity of 40$fT/Hz^{1/2}$@30Hz has been achieved.

\section*{Introduction}
Due to its high sensitivity, SERF magnetometer finds a wide range of application in fundamental physics study and applied instruments. The application includes magnetoencephalography(MEG)\cite{boto2018,borna2017,alemorang2017} and magnetocardiography(MCG)\cite{2019MCGkim,2019mcgwisconsin,Wyllie_2012},rotation sensing\cite{yaochen2016spin,Kornack:2005,jiang2018parametrically}, testing physics beyond the standard model: anomalous spin forces detection\cite{jiwei2018,Terrano:2015,Tullney:2013}, low field NMR detection in a micro-fluid chip\cite{jimenez2014},etc.

The combination of atomic magnetometer with micro-machining technology could fabricate chip-scale atomic magnetometers\cite{alemorang2017,shah2007}. With the merits of small size and low costs, chip-scale atomic magnetometer may find a wide range of application in the industry. In the MEG study, smaller size means more detector could be fixed around the head and this could give more precise source location of magnetic field produced by a bunch of neurons. In electrophysiology, better source location could give a better resolution of abnormal discharge area image before epilepsy surgery. Better resolution of magnetic field  could be very important for the application of atomic magnetometer to Brain-Machine Interface. 

There are basicly two methods to fabricate an atomic magnetometer. One is the traditional machining technology in which the alkali vapor cell is made of glass and the glass vapor cell is handled through a torch. The other method is the micro-machining method in which the vapor cell is fabricated through glass-silicon anode bonding. It is obvious that the bonding technology can fabricate smaller vapor cell and the cost is lower. The glass-silicon bonding cell only owns two transparency glass window sides. The traditional pump-probe configuration is not suitable for the atomic magnetometer. Hence an alkali atom spin modulation method with absorption detection method was developed to configure a single beam SERF magnetometer\cite{shah2007}. 

Even though the simplified configuration only need one beam, the sensitivity of the single beam SERF magnetometer is far less sensitive than that of a pump-probe configured magnetometer. The improvement the sensitivity of the single beam absorption detection is important for chip-scale atomic magnetometer. It is reported that the main noise source of the single beam magnetometer comes from the laser power noise which is originate from a very large background offset after the laser pass through the vapor cell\cite{shah2007}. Here we developed a laser power differential method to suppress the background offset and laser power noise simultaneously to improve the sensitivity of the single beam absorption SERF magnetometer. We achieved a factor of 2 improvement of the laser power noise suppression.

Alkali metal Rb is utilized in most of the SERF atomic magnetometer. The working temperature of Rb magnetometer is around 423K at which Rb atom could react with typical utilized boro-silicate glass in anode bonding. Moreover, Rb owns 2 isotopes which are approximately at the same level in abundance in nature. It is expensive to isolate the two isotopes. Compared with Rb atoms, the working temperature of a Cs SERF magnetometer is around 393K. The heated power consumption is lower than that of Rb. Cs atoms are more difficult to react with glass at the same temperature than that of Rb. At the Cs working temperature, surface coatings such as Octadecyltrichlorosilane(OTS) could withstand such high temperature. The wall relaxation which is much more obvious in micro-machined vapor cells could be sufficiently suppressed in the magnetometers with  OTS surface coatings. At last, there is only one isotope in the nature abundance Cs metal. The cost of alkali vapor cell could be reduced if Cs atom spins are utilized. Hench, we will focus on studying Cs based single beam SERF magnetometer.

In a micro-machined SERF magnetometer, the size of an alkali vapor cell is typically smaller than 3mm in the three dimension directions which is much smaller than the optical table magnetometer in which the vapor cell is larger than 1cm. The neglected wall induced relaxation of atom spins is obvious. In order to suppress this relaxation, Nitrogen gas is typical utilized for it owns a large molecular diameter. However, the spin destruction rate between alkali atom and N$_2$ is large. Thus, in this article, we chose Ne gas as the buffer gas because the spin destruction cross section is 30 times smaller than that of the N$_2$ gas. Meanwhile, the diffusion constant of Cs in Ne is only twice as that of the nitrogen gas. Compared with helium gas which needs special glass to stop its leakage, the leakage rate of Ne would be smaller than that of helium gas for the diameter of Ne molecular is larger. Thus we will focus on studying Cs-Ne SERF magnetometer and we filled 3 amagats Ne in the vapor cell to reduce the wall relaxation in the Cs-Ne magnetometer.
\section*{Theory}
Since it is reported that the main noise source of a single beam absorption SERF magnetometer is from the laser power fluctuation\cite{shah2007}, the zero field magnetic resonance lines could be narrowed to increase the sensitivity of the magnetometer. The line-width of the magnetometer is directly decided by the transverse relaxation of the atomic spins, thus we need to reduce the spin relaxation. In a SERF magnetometer, the electron spins are under spin exchange relaxation free regime. The main relaxation comes from the wall relaxation and the spin destruction relaxation. The basic operation of a SERF magnetometer could be found in this reference\cite{allred2002}. In a large spherical vapor cell whose diameter is 2.5cm, the wall relaxation can be neglected if 1 Amagat of buffer gas such as N$_2$, He and Ne is filled.  In this paper, we focus on micro-machined alkali vapor cell and the cell dimension is smaller than 3mm$\times$3mm$\times$3mm. The relaxation of wall could not be neglected and it could be the main relaxation source.

We can fill higher pressure buffer gas to reduce the wall relaxation rate. However, high gas pressure could also cause the spin destruction relaxation of the atom spins. The trade off between wall relaxation and spin destruction relaxation could be optimized through gas pressure. The utilization of N$_2$ is typical in a micro-machined vapor cell for the diffusion constant of alkali metal in nitrogen is large\cite{shah2007}. However, large nitrogen gas number density could cause substantially spin destruction relaxation. We find that Ne is better than N$_2$ and we have done a simulation to decide which is the better choise for micro-machined vapor cells. 

 Suppose that the laser beam is at the center of the cell and the beam diameter is 0.8mm. The dimension of the cylindrical alkali vapor cell which is the typical micro-machined cell's size is $\phi3mm\times3mm$.   The theory of wall relaxation in a cylindrical cell could be found in this paper\cite{franzen1959}. As the atomic spins moved to the surface of the vapor cell, their spin directions are totally randomized. The diffusion wall relaxation time $T_2^D$ is:
 \begin{center}
$ 1 \div T_2^D = q(\mu^2+\nu^2)D$
\end{center}

where we have set $\mu a=2.405$ and $a$ is the diameter of the cylindrical vapor cell. $\nu=\pi/L$ and $L$ is the length of the vapor cell. The wall relaxation is enhanced by a factor of q because both of the electron and nuclear spin polarization is destroyed at the wall. 

\begin{figure}

\includegraphics[width=9cm,height=7cm]{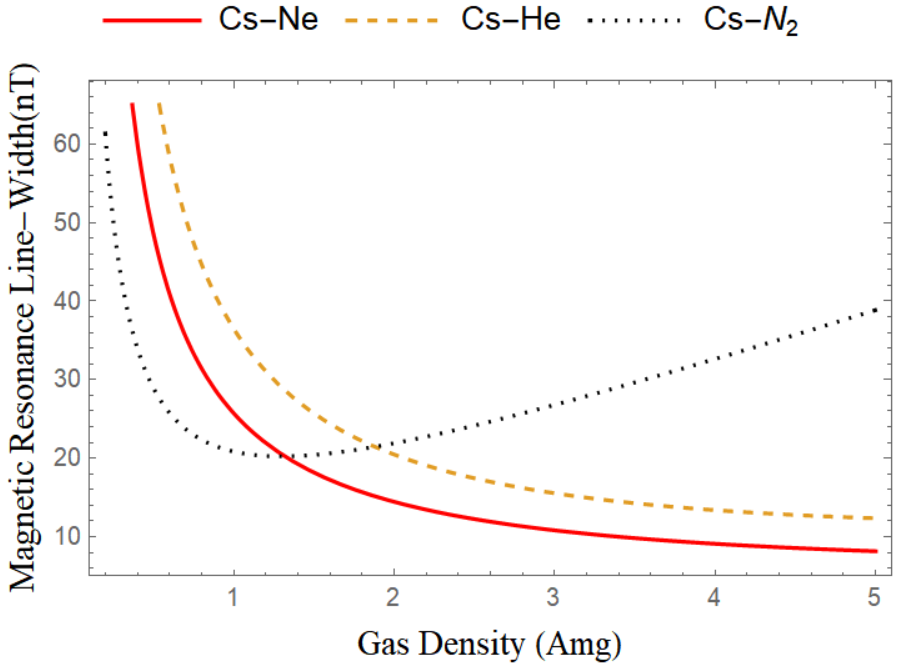}
\captionsetup{labelformat=empty} 
\caption{Figure \ref{fig:csnewall}. The equivalence magnetic resonance line-width resulted from wall relaxation and spin destruction relaxation for Cs atoms at different buffer gas densities.} 
\label{fig:csnewall} 
\end{figure} 

Note that we could not find an experimental parameters for the Cs-Ne spin destruction cross section, instead, we use a theoretical parameters of Cs-$^{21}$Ne spin destruction cross section from this paper\cite{walker1989}. The diffusion coefficient of Cs in Neon gas can be find in this paper\cite{giel2000}. According to the theory analysis of a SERF atomic magnetometer, total relaxation can be divided by the gyro-magnetic ratio of the electron and $2\pi$ to get magnetic resonance line-widths in the unit of nT\cite{yaochen2016spin}. Fig.~\ref{fig:csnewall} shows the relationship between the buffer gas density and the equivalent magnetic resonance line-width for several alkali metal-buffer gas pairs. From the results we conclude that N$_2$ owns a large diffusion coefficient which could efficiently stop the spins from wall spin relaxation. However, with the increasing of buffer gas density, the spin destruction relaxation becomes dominant. Compared with N$_2$, Ne owns a smaller diffusion coefficient. However, the spin destruction relaxation rate is much smaller than that of N$_2$ at high gas pressure. For Ne, we find that the pressure is higher, the line-width is better. We chose the buffer gas pressure to be around 3 amagats which could be easily reached when fabricating an alkali vapor cell. Above this pressure, it is not efficient to reduce the line-width. For Ne gas, the magnetic resonance line-width is 12nT. However, for N$_2$, the best pressure is around 1 amagat and the magnetic resonance line-width is 20nT. therefore we can conclude that Ne is a better candidate for smaller alkali vapor cells. Hence we fill 3 amagats Ne in the alkali vapor cell described in this paper.

The theory of a single beam SERF magnetometer could be found in several articles. The pump-probe configuration is utilized in a traditional SERF magnetometer. In a single beam SERF magnetometer, the absorption of the pump laser is detected to measure the external magnetic field. The absorbed laser power is proportional to the electron spin polarization in the laser propagating direction. Since the magnetometer is only sensitive to the polarization in the transverse plane which is perpendicular to the pumping light direction. A magnetic field modulation technique is utilized to increase the sensitivity of the magnetometer as well as to reduce low frequency laser power noise. The theory and optimization method of the modulation could be found in several references\cite{sheng2017,shah2009}.

\subsection*{Experimental Setup} 
\begin{figure}
\includegraphics[width=9cm,height=6cm]{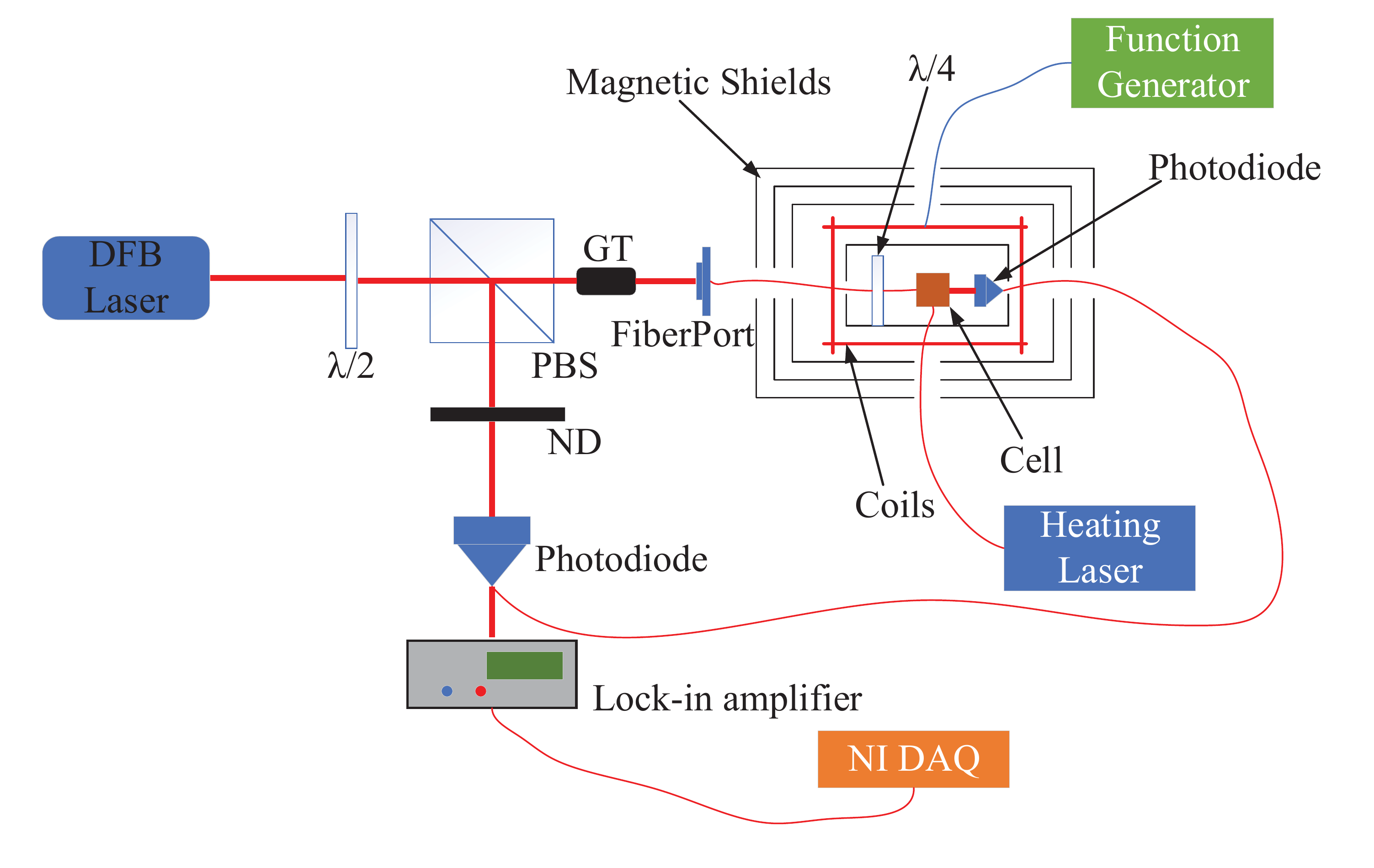}
\caption{Figure \ref{fig:experiment}. The experimental setup of the Cs-Ne SERF magnetometer. PBS:Polarization Beam Splitter; GT:Glan-Thompson Polarizer; ND:Neutral Density Filter; DFB:Distribution Feedback Brag.} 
\label{fig:experiment} 
\end{figure}
\begin{figure}
\includegraphics[width=9cm,height=4cm]{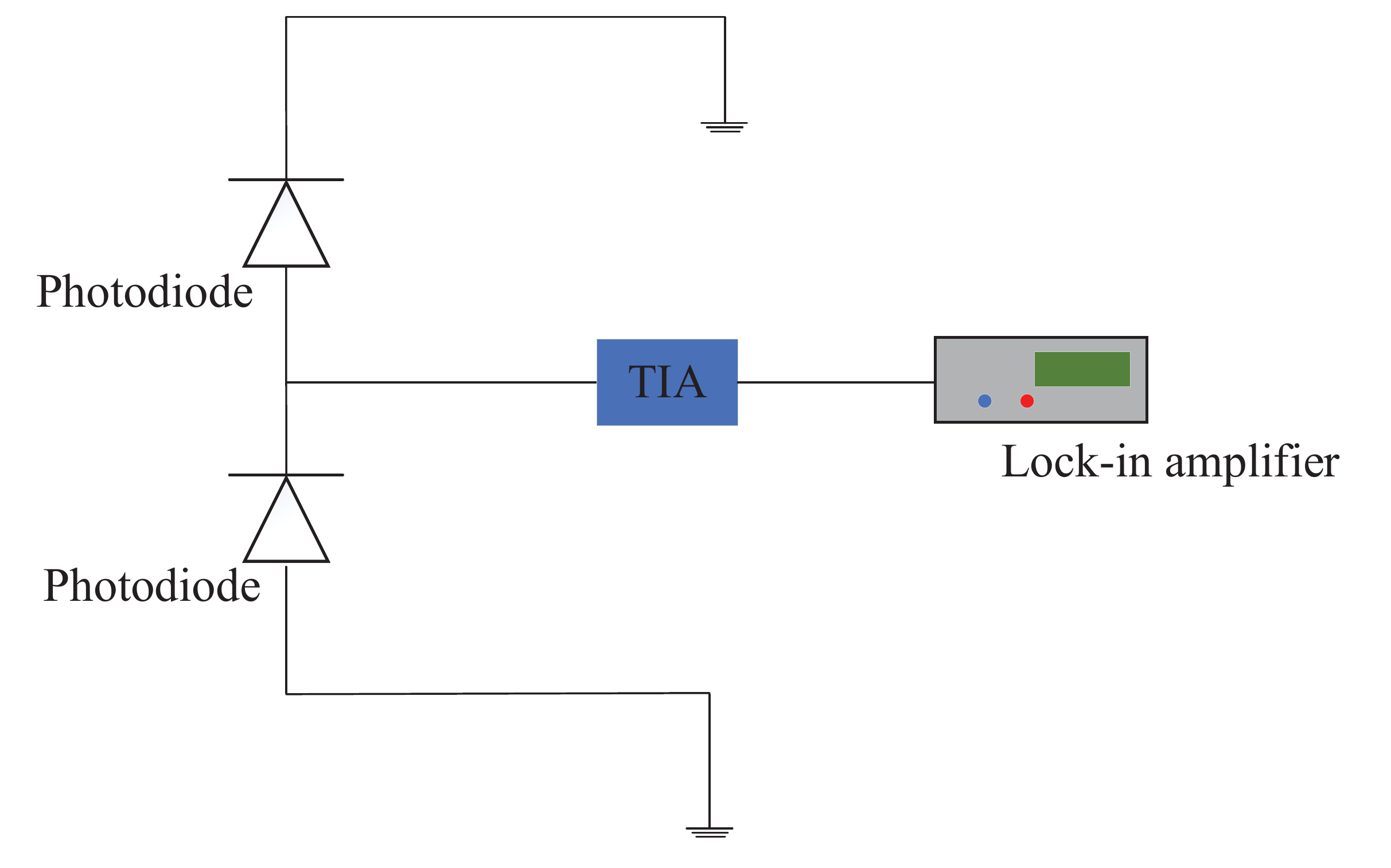}
\caption{Figure \ref{fig:diff}. The principal of the differential circruit for laser background and noise suppression.TIA:Trans-impedence Amplifier.} 
\label{fig:diff} 
\end{figure}

 
The basic configuration of the experiment is shown in Fig.~\ref{fig:experiment}. The cylindrical vapor cell whose inner diameter is 3mm and length is 3mm is made from boro-silicate glass. It contains a small droplet of Cs metal and 3 amagats of Ne gas is filled into it to reduce the wall collisions relaxation and 50 Torr of nitrogen is filled to quench the random photons from decay of the excited state Cs atoms. The cell is heated to 393K through a 3 Watts 1550 nm laser which could efficiently heat the cell as well as produces no magnetic field. The cell is fixed in a PEEK oven for thermal insulation. To eliminate earth magnetic field and its fluctuation, 3 layers of $\mu$-metal shields together with a cylindrical Mn-Zn ferrite are utilized.  The residual magnetic fields could be further canceled actively by a set of homemade tri-axial high precision coils inside the magnetic shields. Besides, the coil is also utilized to modulate the magnetic field to increase the sensitivity of the single beam absorption SERF magnetometer.
In this system, only one DFB laser which is tuned to the Cs D1 line (895nm) and directed along the z axis is used. The $\lambda/2$ wave plate and PBS are utilized to divide the beam into two parts, one beam is set as a reference beam and the other beam passes through the vapor cell for the measurement. The GT polarizer is utilized to further purify the linearly polarized laser light. One polarization maintain fiber is utilized to direct laser light to the cell inside the magnetic field shields. A $\lambda/4$ wave plate changes the laser beam to be circularly polarized. After the cell, a photodiode receives the passed beam. For the reference beam, a neutral density filter is utilized reduce the laser power before it is received by the other photodiode. The differential circruit showed in Fig.~\ref{fig:diff} is composed of the two photodiodes. The current could be substracted if the laser power of the two beam is equal. The circruit could suppress the background offset and laser power noise simultaneously to improve the sensitivity of the SERF magnetometer. The differential signal is amplified through a TIA and then it is demodulated by a lock-in amplifier such that the output of the magnetometer is proportional to the input magnetic field\cite{shah2009}. The response of the magnetometer to the magnetic field is perpendicular to the direction of the pump beam. The magnetic field signal finally was acquired through a NI(National Instrument) DAQ(data acquisition card). 
\section*{Results}
The magnetic resonance line-width had been simulated in the last section. In this section, we first do a measurement of the line-width. This measurement not only gives information about the total relaxation of the magnetometer, but also can be utilized to optimize the magnetometer to a better laser power working point. The temperature of the vapor cell is heated to 385K. Since 3 amagats of Ne are filled in the vapor cell. The broadened laser absorption line-width is 27.6GHz. These parameters will give an optical depth of 2.2. A modulation magnetic field with frequency of 990Hz is added to the x direction. The modulation amplitude is optimized. At certain optical power, the modulation amplitude is changed and the response of the magnetometer to DC x magnetic field is recorded. When the response is the largest, the modulation amplitude is the optimized one.

\begin{figure}
\includegraphics[width=9cm,height=4cm]{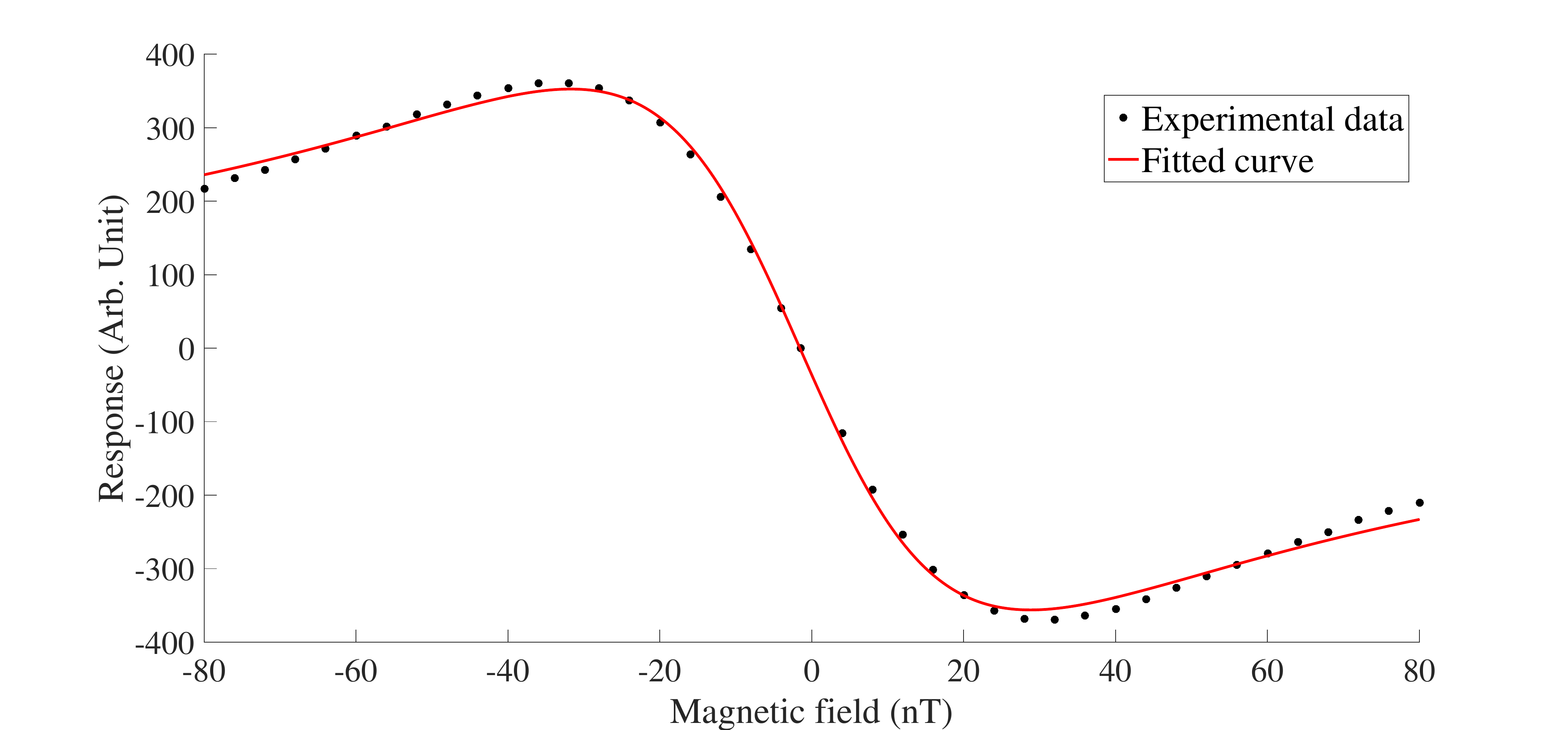}
\caption{Figure \ref{fig:response}. The response of the magnetometer to a static x transverse magnetic field.} 
\label{fig:response} 
\end{figure} 
We found that the optimized amplitude is 200nT peak to peak. After the modulation amplitude optimization, we did a zero field magnetic field resonance line shown in Fig.~\ref{fig:response}\cite{Ledbetter2008}. We change the transverse x magnetic field and the response of the magnetometer is recorded. A Lorentz curve is fitted and we can get the line-width of the curve to be 30nT. This line-width equals to a total relaxation of 5275 sec$^{-1}$, including the pumping rate, wall relaxation, Cs-Ne spin destruction relaxation and Cs-Cs spin destruction relaxation. We changed the power of the laser then recorded several line-widths at different laser power. Then we extrapolate the line-width to zero power point and we get a line-width of 16nT. Since the magnetometer responses largest if the electron polarization is 50$\%$, we need to set the laser power of the magnetometer to a point when the line-width is 32nT.  At the best laser power point, we did several magnetic field sensitivity measurements.

\begin{figure}
\includegraphics[width=9cm,height=5cm]{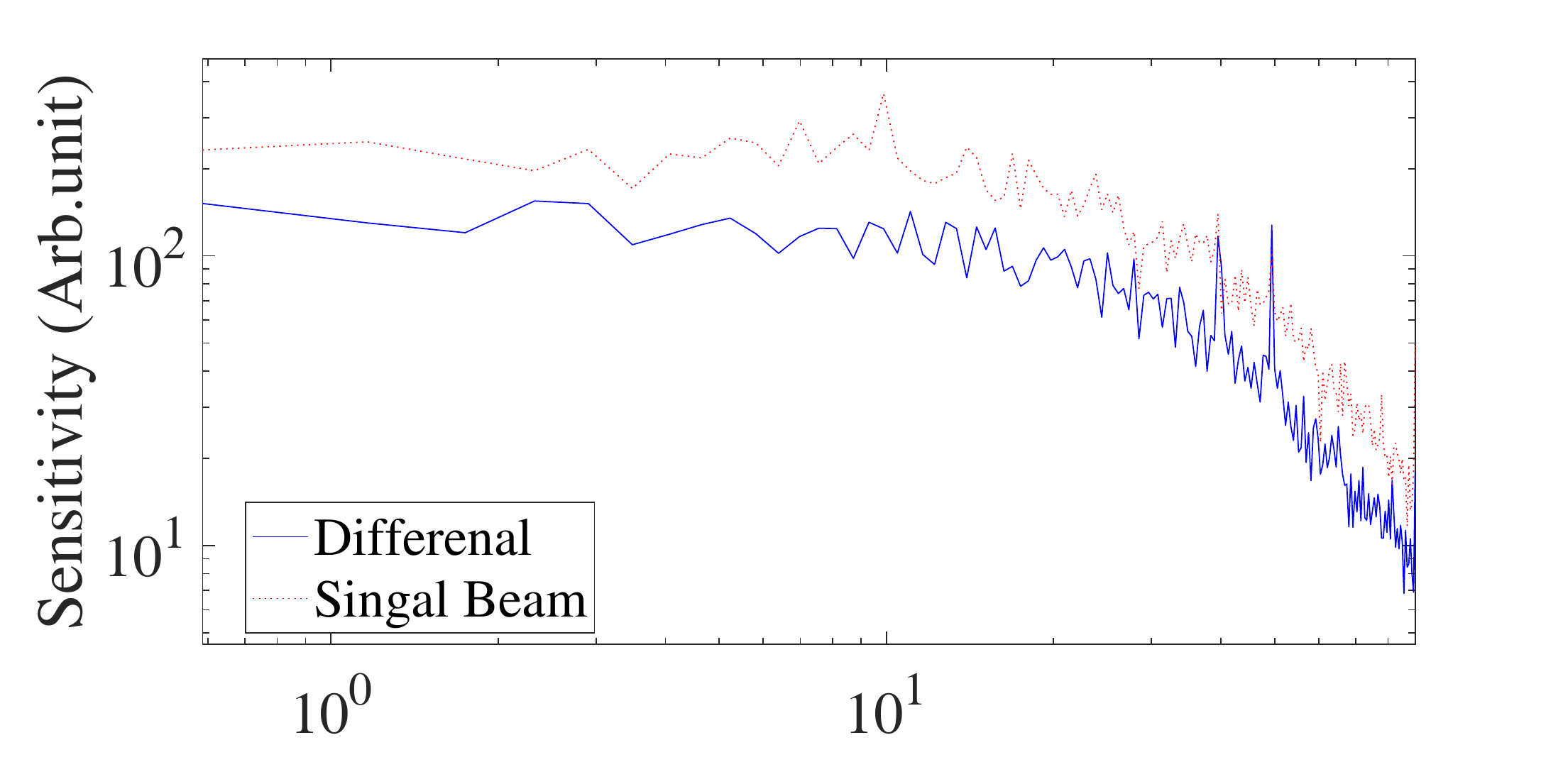}
\caption{Figure \ref{fig:differential}. The noise spectrums of the laser power for the single beam and the differential signal.} 
\label{fig:differential} 
\end{figure} 

Since the absorption of the pumping laser is detected in the single beam SERF magnetometer, there is a background offset in the detected signal. The reference laser is utilized to do a differential measurement to suppress the background offset. Moreover, the laser power noise at the modulation frequency can also be suppressed in this differential method. The diameter of the laser in our experiment is 0.8mm. We adjust the reference laser power attenuation and the transmitted 22$\mu$A background current could be suppressed to be smaller than 500nA. In order to evaluate the differential method, we measured the noise spectrums of the laser power with the differential method and without the differential method. The red dotted line in Fig.~\ref{fig:differential} shows the noise spectrum of the laser power noise without the differential method. The blue solid line is the noise spectrum of the laser power with differential method. Note that the signals are taken when the frequency of the laser is far from the absorption center and there is nearly no interaction between the laser light and the atoms. As shown in Fig.~\ref{fig:differential}, there is a factor of 2 improvement at the range of the magnetometer working frequency range.The laser power noise at 990Hz is suppressed by a factor of 2 in our experiment setup. 

\begin{figure}
\includegraphics[width=9cm,height=6cm]{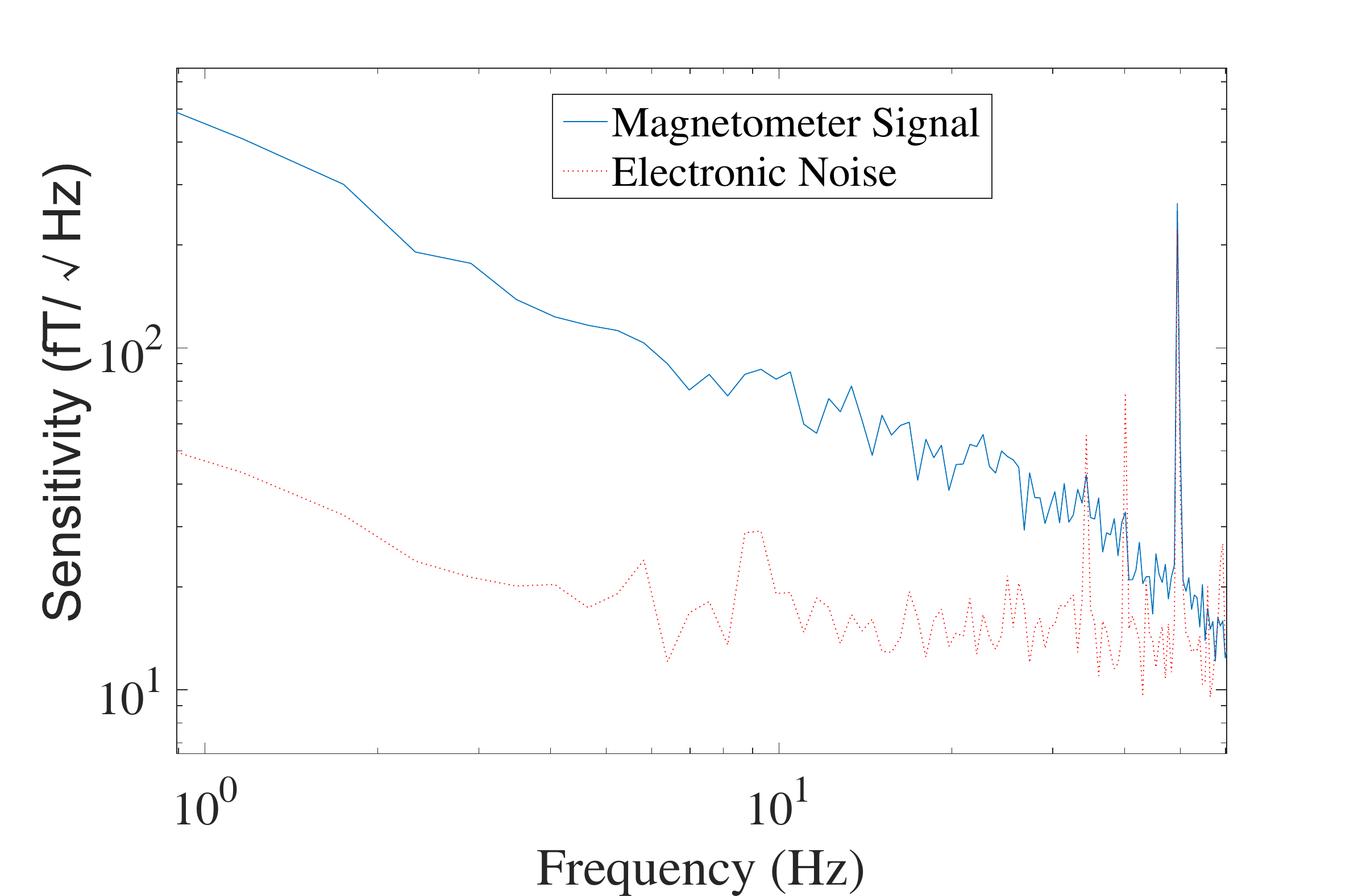}
\caption{Figure \ref{fig:sensitivity}. The sensitivity of the magnetometer.} 
\label{fig:sensitivity} 
\end{figure}

After optimization of the laser power and the modulation amplitude, we measured the sensitivity of our magnetometer. First we need to acquire the scale factor. We give a standard DC magnetic field in the x direction, then the output voltage change is acquired. In order to get the sensitivity, we adjust the DC input magnetic field to 0, then the voltage signal is recorded. We did a power density analyse of the noise signal. In order to give a more accurate result, we averaged the noise spectrum in the 1Hz range. Fig. ~\ref{fig:sensitivity} shows the spectrum of the magnetic field noise. The resolution of the frequency axis is 1Hz. We also shut off the laser then did a similar spectrum analyse for the electronics and the photo-diode dark current noise. Note that the scale factor is frequency depended because our magnetometer owns a finite bandwidth. The frequency response of the SERF magnetometer is like an one order low pass filter\cite{kominis2003}. The cutoff frequency\cite{allred2002} could be calculated through the spin relaxation rate. In our experiment, the zero magnetic resonance line-width is 30nT which corresponds to a total spin relaxation of 5275 sec$^{-1}$. Together with the polarization of 0.5 for Cs atom spins, the cutoff frequency is calculated to be 516Hz. From Fig.~\ref{fig:sensitivity} we can get that the sensitivity of our magnetometer is 40$fT/Hz^{1/2}$@30Hz.
\section*{Conclusion}
In conclusion, we have developed a Cs-Ne SERF magnetometer which is specially designed for the future atomic magnetometer chip with micro-machined fabricating method. We compared nitrogen, helium and neon buffer gas. The neon gas is better than nitrogen which is typical utilized in current chip-scale atomic magnetometer. Since the silicon-glass bonding vapor cell only owns two transparent windows, the single beam absorption spin detection method is typical utilized. In order to reduced the large background offset and the laser power noise. A differential method is developed and a factor of 2 improvement could be realized. We finally get a magnetic field sensitivity of 40$fT/Hz^{1/2}$@30Hz. Compared with Rb atoms, Cs atoms work at a lower temperature. The typical boro-scilicate glass which is utilized in the anode bonding process could withstand the chemical corrosion of Cs atoms. The power consumption of the magnetometer should also be smaller.
\section*{Acknowledgments}
This work is supported by Zhejiang Lab under grant number 2019MB0AB02, 2019MB0AE01 and 113009-AA2003, China Postdoctoral Science Foundation under grant number 2020M683462 and 2019M662121 and Jiangsu Province Youth Foundation under grant number BK20200244.


\end{document}